\documentclass[useAMS,usenatbib]{mn2e}
\usepackage{graphicx}
\newcommand {\be}{\begin{equation}} 
\newcommand{\ee}{\end{equation}}    
\newcommand{\sss}{\scriptscriptstyle}
\def\nabp{\nabla_{\perp}}
\def\ddt{\frac{\partial}{\partial t}}

\def\vti{v_{{\sss T}i}}
\def\vtiz{v_{{\sss T}iz}}
\def\vte{v_{{\sss T}e}}

\title[The universally growing mode in the solar atmosphere:
coronal heating by drift waves]{The universally growing mode in the
solar atmosphere: coronal heating by drift waves}
\author[J. Vranjes and S. Poedts]{J. Vranjes\thanks{E-mail:
Jovo.Vranjes@wis.kuleuven.be; jvranjes@yahoo.com} and S.
Poedts\thanks{E-mail:
Stefaan.Poedts@wis.kuleuven.be}\\
K. U. Leuven, Center for Plasma Astrophysics, Celestijnenlaan 200B,
3001 Leuven,
 Belgium,\\ and Leuven Mathematical Modeling and Computational Science Center
 (LMCC)}
\begin{document}

\date{Accepted xxx. Received xxx; in original form xxx}

\pagerange{\pageref{firstpage}--\pageref{lastpage}} \pubyear{2002}

\maketitle

\label{firstpage}

\begin{abstract}
The heating of the plasma in the solar atmosphere is discussed
within both frameworks of fluid and kinetic drift wave theory.  We
show that the basic  ingredient necessary for the heating is the
presence of density gradients in the direction perpendicular to the
magnetic field vector. Such density gradients are a source of free
energy for the excitation of drift waves. We use only well
established basic theory, verified experimentally in laboratory
plasmas. Two mechanisms of the energy exchange and heating are shown
to take place simultaneously: one due to the Landau effect in the
direction parallel to the magnetic field, and another one,
stochastic heating, in the perpendicular direction. The stochastic
heating i)~is due to the electrostatic nature of the waves, ii)~is
more effective on ions than on electrons, iii)~acts predominantly in
the perpendicular direction, iv)~heats heavy ions more efficiently
than lighter ions, and v)~may easily provide a drift wave  heating
rate that is orders of magnitude  above  the value that is presently
believed to be sufficient for the coronal heating, i.e., $\simeq 6
\cdot 10^{-5}\;$J/(m$^3$s) for active regions and $\simeq 8 \cdot
10^{-6}\;$J/(m$^3$s) for coronal holes. This heating acts naturally
through  well known effects that are, however, beyond the current
standard models and theories.

\end{abstract}


\section{Introduction}

The physical parameters in all three main regions of the solar
atmosphere (photosphere, chromosphere and corona) change both
horizontally and with altitude. This by all means also includes the
collisional frequency, with respect to which the solar atmosphere
could be, roughly speaking, termed as `strongly collisional' (in the
photosphere and in the lower part of the chromosphere), `mildly
collisional' (in the upper part of the chromosphere), and
`collision-less' (in most of the corona and beyond). This
classification is not absolute as it depends on the ratio of the
characteristic time $\tau_c$ for certain physical process, and the
collisional time $\tau_\nu \sim1/\nu$, where $\nu$ is the collision
frequency.  The collisions of plasma species include those between
plasma particles (e-i, e-e, i-i), i.e., Coulomb type collisions
(which are typical for the corona), and those between plasma
particles and un-ionized neutrals, e-n and i-n (which are absolutely
dominant in the photosphere and in the lower part of the
chromosphere). In the photosphere, for example, the ion-neutral and
electron-neutral collision frequencies \citep{v1,v2,pand} are about
$10^9$ and $10^{10}\;$Hz, respectively, while the ion-ion and
electron-ion Coulomb collision frequencies are around  $10^7$ and
$10^{9}\;$Hz, respectively. In the corona, the Coulomb collisions
occur much less frequent: $\nu_{ei}\simeq 36\;$Hz and
$\nu_{ii}\simeq 1\;$Hz (for $n_0=10^{15}\;$m$^{-1/3}$), and
$\nu_{ei}\simeq 0.4\;$Hz and $\nu_{ii}\simeq 0.01\;$Hz (for
$n_0=10^{13}\;$m$^{-1/3}$).

All plasma modes (with probably  one exception only, see below) are
generally damped by collisions, and therefore a strong source is
needed whenever one deals with waves in the (lower) solar
atmosphere. However, such a (local!) source sustaining the waves
over longer distances or time intervals is most often absent, even
in models for coronal heating by waves. On the other hand, in the
practically collision-less and very hot corona, Landau damping
occurs and damps plasma waves very effectively by hot resonant ions,
so that in this region too, a wave source is required to sustain the
waves over a longer time interval.

There is only one mode that is able to survive the drastically
different (collisional-collisionless) extremes in the different
layers of the solar atmosphere, viz.\ {\em the drift mode}.
Moreover, this mode is able to benefit (grow) from each of these
extreme situations. This drift mode has been called the `universally
growing mode' in the literature. In collisional plasma, the drift
mode grows due to the electron collisions and this can be described
within the two-fluid model. In collision-less plasma, however, the
mode grows due to the electron resonance effect in the presence of a
density gradient, but this is a purely kinetic effect. The fluid
description works very well in the lower solar atmosphere, simply
due to the fact that the plasma is so strongly collisional there
(i.e., the maxwellization is very effective). On the other hand, the
kinetic description is preferred in the collision-less environment
in the corona. Hence, in both extremes we have a proper environment
and a proper theoretical model for studying the strong instabilities
of the drift waves. Nevertheless, the driving mechanism for the
waves is the same in both cases, namely {\em the presence of a
density gradient} perpendicular to the ambient magnetic field
vector.

Numerous observations confirm the presence of such density
irregularities. Yet, the impression is that the possible role of the
drift wave in the coronal heating problem is either overlooked or
strongly underestimated (the mode is never mentioned in any book
dealing with solar plasma). This may be partly due to the fact that
the drift wave necessarily implies a multi-component fluid or
kinetic description, contrary to the widely used single-fluid
magnetohydrodynamics (MHD) model, within which the drift wave simply
can not be studied.

\begin{figure}
  \includegraphics[height=8cm,bb=55 32 533 577,clip=,width=.85\columnwidth]{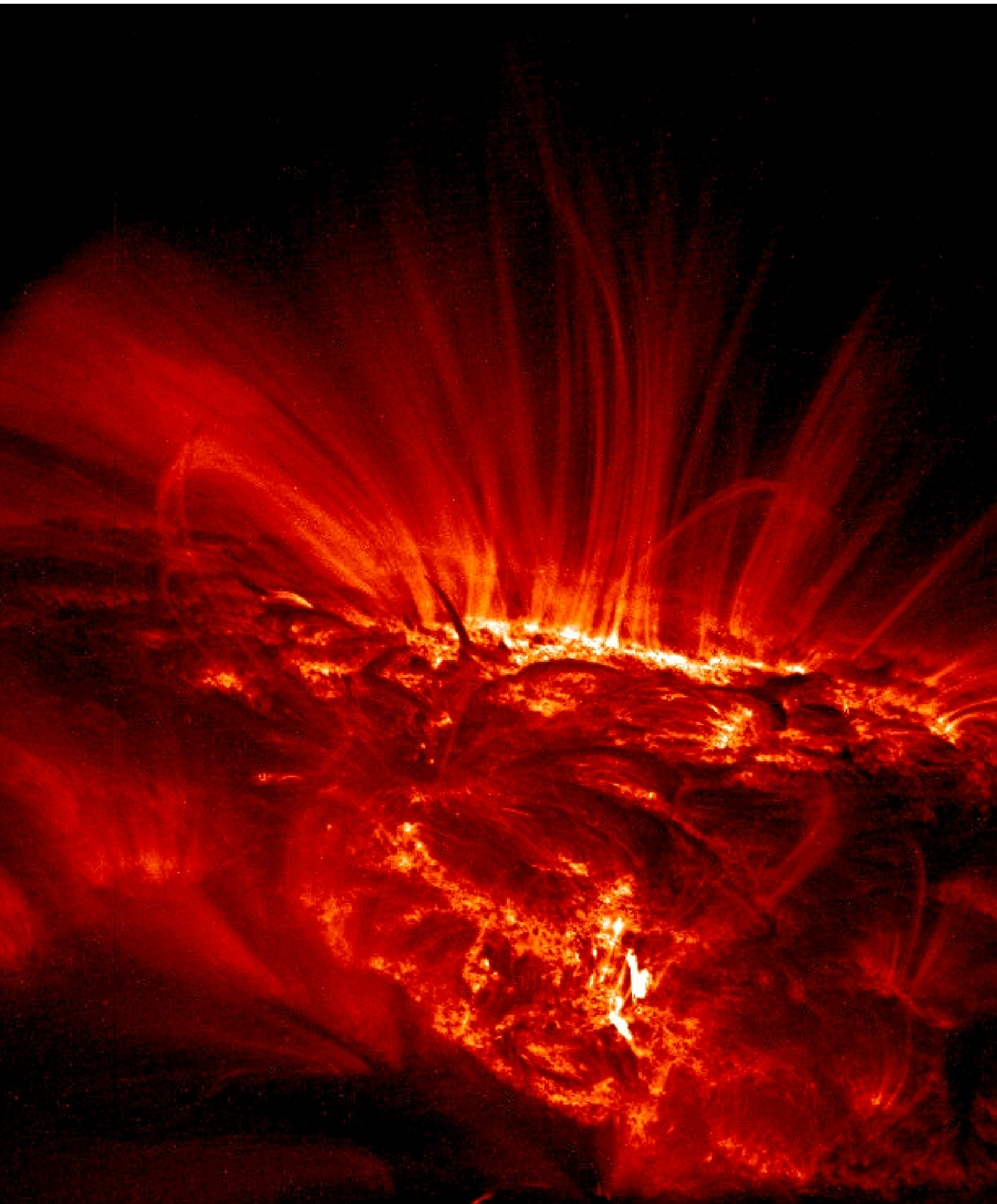}\\
   \includegraphics[height=8cm,bb=32 33 757 577,clip=,width=.85\columnwidth]{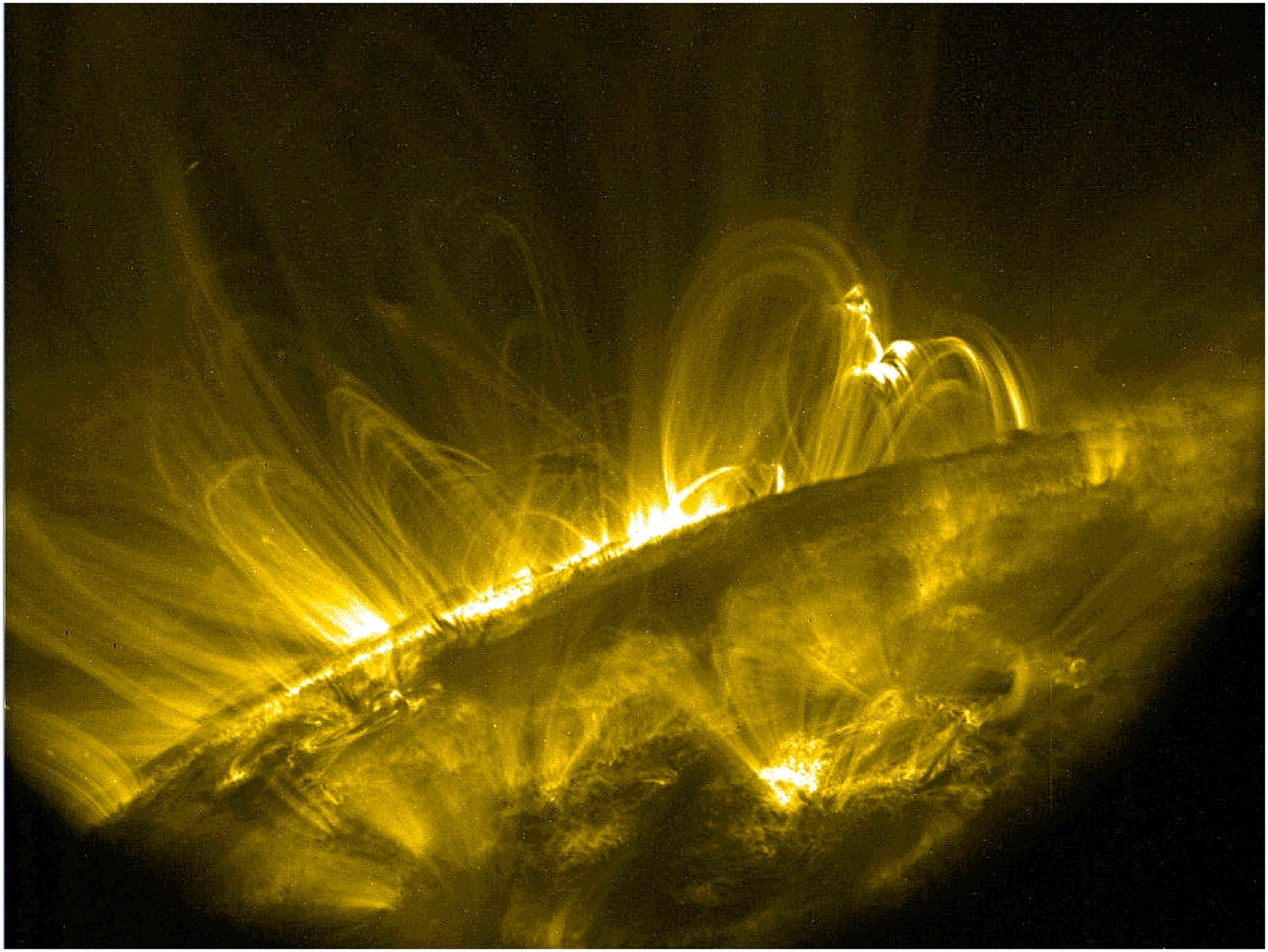}\\
\vspace*{2mm}
 \caption{TRACE photos of numerous loops in  active
regions, with clear density structures.} \label{fig1}
\end{figure}

The heating of the corona is one of the long-standing puzzles in
solar physics and relates to the question of why the temperature of
the Sun's corona is about 1 to $3\times10^6\;$K (parts of the corona
are even hotter) while the solar surface is only around $5600\;$K
hot. A detailed review of the problem may be found in
\citet{nar,klim}. According to \citet{nar} the necessary heating
rates for active regions and coronal holes are around $6 \cdot
10^{-5}\;$J/(m$^3$s) and  $8 \cdot 10^{-6}\;$J/(m$^3$s),
respectively. Similar values and a detailed analysis of the problem
may be found also in \citet{asb}. Two theories on coronal heating,
namely wave heating [e.g. \citet{suz}] and magnetic reconnection (or
nanoflares), have remained as the most likely candidates, and many
solar physicists believe that some combination of these two theories
can probably explain coronal heating, although `the details are not
yet complete'.   Both models, however, rely on the continuum
approximation (MHD) while it is clear that the actual heating takes
place at length scales much smaller than those on which the MHD
model is justified. Moreover, it is evident that the observed
discrepancy between the ion and electron temperatures in the corona
as well as the observed large temperature anisotropy, with a proton
perpendicular temperature higher than the parallel temperature, are
beyond the (single!) fluid model. Note also that according to a
recent study \citep{reg}, the distribution of magnetic null points
(only 2\% of them are located in the corona and 54\% in the
photosphere) is opposite to what would be required for the mechanism
that is supposed to heat the corona. The heating by waves rather
than by reconnection is also supported by the diagnostic of active
regions presented in \citet{mil}.

A self-consistent heating model must fulfil a lot of requirements.
It must: 1)~provide an energy source for the  extremely high
temperature in corona, including 2)~a reliable and efficient
mechanism for the energy transfer from the source to the plasma
particles, and 3)~this with a required heating rate. It should also
4)~explain the discrepancy between ion and electron temperatures
(typically $T_i>T_e$), 5)~explain the origin of the large
temperature anisotropy ($T_{\bot} > T_{\|}$) with respect to the
direction of the magnetic field, particularly for ions, 6)~explain
the observed larger heating of heavier ions, and last but not least,
7)~it should work everywhere in corona (with well known different
heating requirements in active and quiet regions).

In this paper, we present a new coronal heating model that can
operate in all layers and in all magnetic structures of the solar
atmosphere and that is able to explain all seven  requirements given
above. Our model  represents a  {\em new paradigm} in which a)~the
energy for driving the drift modes and for the heating of corona is
already present in corona, and, b)~this energy is naturally
transmitted to the different plasma species by well known effects
that are, however, beyond the standardly used models and theories.
Moreover, it is  based on well established, basic theory which has
already been verified and confirmed by means of laboratory plasma
experiments. All that is needed for the heating mechanism to work is
the presence of a density gradient perpendicular to the magnetic
field. In the solar corona, this may be taken rather as a fact than
as a hypothesis.

\section{Inhomogeneous solar plasma}

Fine density filaments and threads  in the solar atmosphere have
been observed for a long time now, even from ground-based
observations like those during the eclipse in 1991 \citep{nov},
showing a slow radial enlargement of plasma structures. On the other
hand, measurements by Voyager~1 and Voyager~2 show  \citep{woo} that
the finest structures in the slow solar wind at around
$9\;R_{\odot}$ are about 3 times finer than those in the fast wind.
Assuming a radial expansion, these authors conclude that the
transverse sizes of these highly elongated structures at the Sun are
below 1$\;$km. The contour maps presented in \citet{kar} reveal the
existence of numerous structures of various size. The smallest
filamentary structures of the order of $1\;$km have been discussed
also in \citet{w}.  Recent Hinode observations \citep{bart} confirm
that the solar atmosphere is a highly structured and inhomogeneous
system  and revealed radially spreading grass-like density filaments
of various size pervading the whole domain. A very recent
three-dimensional analysis \citep{aas} of coronal loops reveals
short-scale density irregularities within each loop separately.
Particularly clear images of coronal loops with density structures
may be seen in \citet{as1}  and also in \citet{var}. In Fig.~1 we
give some TRACE images showing many coronal loops in active regions.
Such active regions do not cover the whole corona, yet the analysis
from \citet{as2} reveals that they require around 82\% of the total
energy needed to heat the entire corona.

The presence of plasma density irregularities throughout the corona
implies a plasma that is not in thermodynamic equilibrium. In other
words, it reveals the presence of free energy in the system.  In all
the examples mentioned above, such density irregularities are as
rule associated with the magnetic field, creating the ideal
environment for drift waves. The purpose of this work is to show
that the energy stored in these density gradients may drive drift
waves on massive scales. We shall point out some basic features of
the drift wave instability, apply them throughout the solar
atmospheric plasma and compute the appropriate growth rates. It will
be shown that the growing drift wave and its subsequent interaction
with plasma particles may yield the long searched solution of the
problem of the heating of the solar atmosphere. The model implies
that the direct energy supply for the heating comes from within the
corona itself, though still maintained and replenished by some
mechanisms from below the photosphere. Those include a continuous
restructuring of the magnetic field  implying the consequent similar
changes of the plasma density (due to the frozen-in condition), and
also the observed inflow of the plasma along the magnetic loops
\citep{sc}. To some extent, this looks similar to the currently
accepted scenarios mentioned above, where  the magnetic field plays
an essential role and is  assumed as given. However, in this new
approach the dissipation of these drift waves is easy to explain in
our kinetic model that works on the (very small) length scales at
which the actual dissipation takes place.

The observable characteristic dimensions of the density
irregularities are limited  by the available resolution of the
instruments (about 0.5 arcsecond in the example from Fig.~1, that is
below $400\;$km on the Sun). Nevertheless, even extremely short,
meter-size scales can not be excluded, especially in corona
\citep{v4}. This can be seen by calculating the perpendicular ion
diffusion coefficient \citep{chen} for a coronal environment:
$D_{\bot, j}\approx \kappa T_j \nu_j/(m_j \Omega_j^2)\propto
m_j^{1/2}$, where $j=i, e$. Taking  $B_0= 10^{-2}\;$T, $n_0=
10^{15}\;$m$^{-3}$, and $T_e=T_i=10^6\;$K, for ions we obtain
$D_{\bot, i}=0.01\;$m$^2$/s. The diffusion velocity in the direction
of the given density gradient is \citep{v3} $D_{\bot,j} \nabla
n_0/n_0$ [see also Eq.~(\ref{e3}) further in the text]. Taking the
inhomogeneity scale-length $L_n\equiv [(dn_0/dx)/n_0]^{-1}=10$,
$10^2$, and $10^5\;$m,  where $x$-denotes the direction
perpendicular to the magnetic field vector, we obtain for the ion
diffusion velocities, respectively, $10^{-3}$, $10^{-4}$, an
$10^{-7}\;$m/s only. Therefore, even very short density
inhomogeneities can last long enough to support relatively high
frequency drift instabilities. Hence, in dealing with the drift
wave, we may operate with  the density inhomogeneity scale lengths
that  have any value ranging from one meter up to thousands of
kilometers in the case of coronal plumes.

\section{Drift  wave within fluid theory}

Assuming a partially/weakly ionized and collisional plasma, like in
the photosphere and chromosphere, it is justified to employ the
fluid model. In such an environment, the kinetic Landau damping is
not expected to play any significant role as long as the ion mean
free path is below the wavelength.  The strong-weak Landau damping
transition has been experimentally verified \citep{da} to be at the
threshold $\omega\sim \nu_i$. The  momentum equations that we use
for electrons and ions can be written as
\[
 m_in_i\left[\frac{\partial \vec v_i}{\partial t} + (\vec
v_i\cdot\nabla) \vec v_i\right] = e n_i \left(-\nabla \phi  + \vec
v_i\times \vec B_0\right)
\]
\be
- \kappa T_i\nabla n_i - \nabla\cdot \pi_i- m_in_i\nu_i \vec v_i,
\label{e1} \ee and
\be
 0 = -e n_i \left(-\nabla \phi  + \vec v_e\times \vec B_0\right)
 - \kappa T_e \nabla n_e  - m_e n_e \nu_e \vec v_e,
 \label{e2} \ee
respectively. Here, $\nu_i\equiv \nu_{in}$ and $\nu_e=\nu_{en}+
\nu_{ei}$.  The ion momentum change due to collisions with electrons
is neglected in view of the mass difference. This is justified as
long as the ion dynamics in the direction of the magnetic field
vector is negligible. Otherwise it may modify the instability
threshold \citep{v6}. The shape of the equations reveals that we are
dealing with electrostatic perturbations, the hot ion effects are
included through the pressure and the {\em gyro-viscosity} stress
tensor terms, while the left-hand side of the electron momentum
equation is omitted, implying perturbations with phase speed and
perturbed velocity both much below the electron thermal velocity.
The dynamics of the neutrals may also be included. However, as shown
elsewhere \citep{v5}, this  usually yields small or negligible
corrections, and  such a  model works well as long as the  ion sound
response is negligible. Otherwise, there is an instability threshold
that is modified when the dynamics of the neutrals is
self-consistently included \citep{v6}. A typical geometry of the
drift wave in cylindric coordinates is presented in Fig.~2. Here, as
an example we consider a wave propagating in the poloidal and axial
directions, with the poloidal mode number $m=2$, in a plasma with a
radially dependent equilibrium density. The actual wave fronts are
twisted around the axis and have an $r$-dependent amplitude which
reaches a maximum in the area of the largest density gradient. The
mode behavior in the presence of both radial and axial equilibrium
density gradients is discussed in \citet{veig}.

\begin{figure}
\includegraphics[height=6cm, bb=-7 -6 840 526, clip=,width=.95\columnwidth]{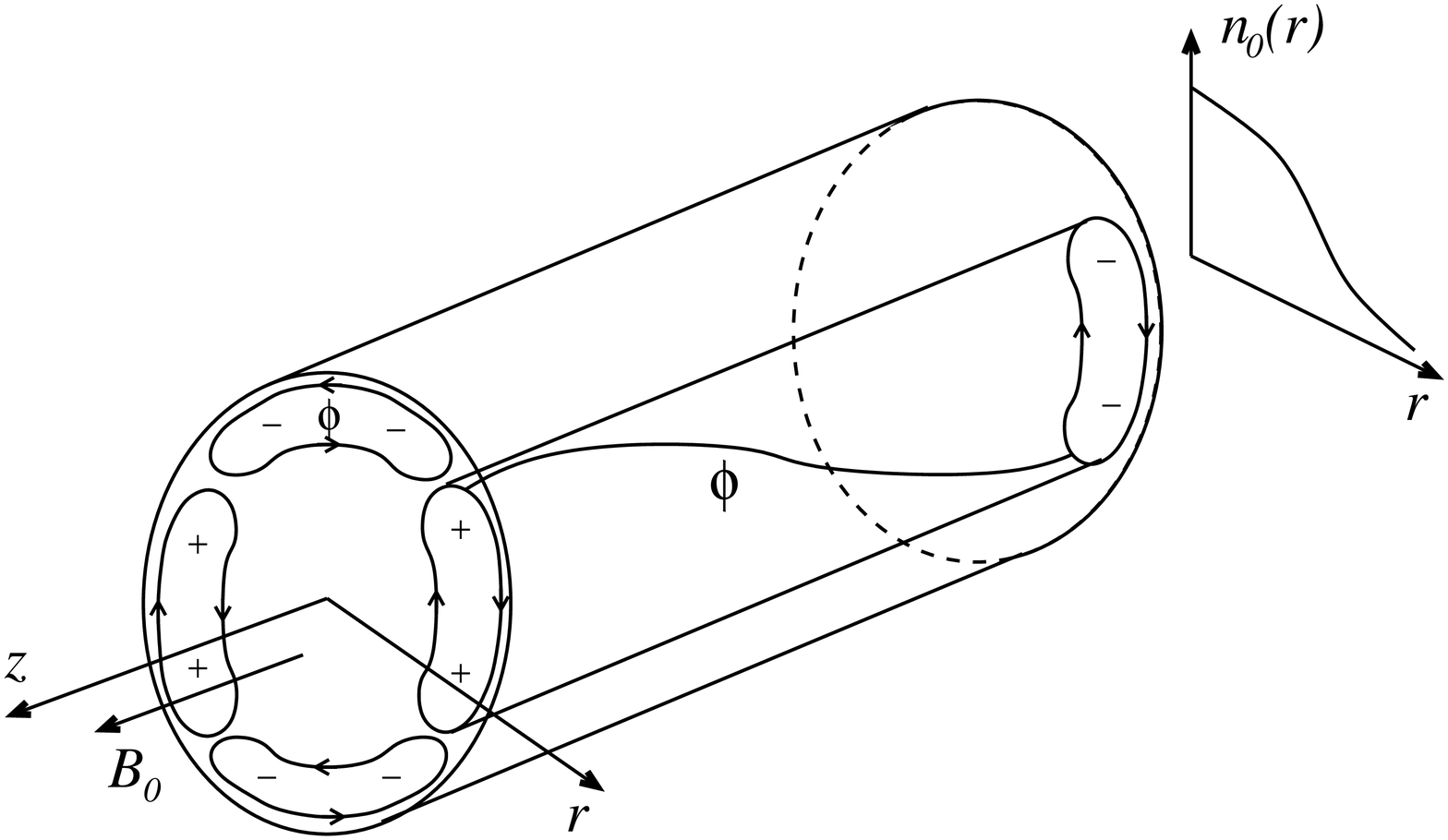}
\vspace*{3mm} \caption{A drift wave  in cylindric geometry with
poloidal wave number $m=2$. }\label{fig2}
\end{figure}
In some regions, the drift mode may become electromagnetic, provided
the plasma $\beta$ exceeds the electrostatic limit. This yields a
coupling of the drift and the kinetic Alfv\'en wave \citep{kad,v3}.
The first {\em experimental observation} of such a coupled mode  in
a hot-ion plasma is discussed in \citet{nis}. The drift-wave part of
the observed coupled modes, is strongly growing, and the maximum
growth rate $\omega_i\simeq 0.1\, \omega_r$, where $\omega_r$ is the
drift wave frequency. In the case of the lower solar atmosphere, as
shown in \citet{v3},  the Alfv\'en part of the mode is of no
interest as it is  always damped due to the collisions, and it will
not be discussed here. Note that the usual gas viscosity is as a
rule negligible even for a relatively high density environment like
the photosphere and chromosphere. More details are available in
\citet{v2}.

The ion gyro-viscous components of the stress tensor that we need
here are \citep{weil}
\[
  \pi_{xy}=\pi_{yx}=\frac{n_i \kappa T_i}{2 \Omega_i} \left(\frac{\partial v_{ix}}{\partial x}
  - \frac{\partial v_{iy}}{\partial y}\right),
  \]
  \[
 \pi_{yy}=-\pi_{xx}=\frac{n_i \kappa T_i}{2 \Omega_i} \left(\frac{\partial v_{iy}}{\partial x}
 + \frac{\partial v_{ix}}{\partial y}\right).
\]
Here and further we have taken $\vec B_0= B_0 \vec e_z$, and
$\vti^2=\kappa T_i/m_i$.

The ion perpendicular velocity  obtained from Eq.~(\ref{e1}) is
described by the following recurrent formula
\[
v_{i\bot}=\alpha_i\left[\frac{1}{B_0} \vec e_z \times \nabp \phi
 + \frac{v_{{\sss T}
i}^2}{\Omega_i} \vec e_z \times \frac{\nabp n_i}{n_i}
-\frac{\nu_i}{\Omega_i} \frac{\nabp \phi}{B_0}\right.
\]
\[
- \left. \frac{\nu_i v_{{\sss T} i}^2}{\Omega_i^2} \frac{\nabp
n_i}{n_i} + \vec e_z \times \frac{\nabp\cdot \pi_i}{m_i n_i
\Omega_i}  + \frac{1}{\Omega_i} \left(\ddt + \vec v_i  \cdot \nabla
\right) \vec e_z \times \vec v_{i \bot} \right.
\]
\be \left. - \frac{\nu_i}{\Omega_i} \frac{\nabp\cdot \pi_i}{e n_i
B_0} - \frac{1}{\Omega_i} \frac{\nu_i}{\Omega_i} \left(\ddt + \vec
v_i \cdot \nabla \right) \vec v_{i \bot}\right]. \label{e3}
 \ee
Here, $ \alpha_i= 1/(1+ \nu_i^2/\Omega_i^2)$. The velocity  can be
calculated up to small terms of any order using the drift
approximation $|\partial/\partial t|\ll \Omega_i$.

\subsection{Ion diamagnetic current effects}

The ion gyro-viscosity is usually overlooked in the literature and,
in particular, in the domain of solar plasma, and this even in
studies using the multi-component plasma theory. For the solar
plasma case, this may have very important consequences because ion
and electron temperatures are typically  of the same order, so that
a cold ion approximation and the consequent neglect of the ion
gyro-viscosity can not  be justified.

As a matter of fact, it is well known \citep{v3,weil}  (yet
standardly disregarded) that part of the ion gyro-viscosity
contributes to the cancelation of some terms in the ion continuity
equation. Clearly, this implies that, if the gyro-viscosity terms
are omitted in the derivation, the resulting ion equations  contain
terms that can not possibly be there, yielding some false  physical
effects. The cancelation of terms appears when Eq.~(\ref{e3}) is set
into the ion continuity $\partial n_i/\partial t+ \nabla_\bot\cdot
(n_i v_{i\bot}) + \nabla_z\cdot(n_i v_{iz})=0$, and it involves,
respectively, the diamagnetic and polarization drifts from one side
\[
\vec v_{*i}=\frac{v_{{\sss T} i}^2}{\Omega_i} \vec e_z \times
\frac{\nabp n_i}{n_i}, \quad \vec v_{pi}=\frac{1}{\Omega_i}
\left(\ddt + \vec v_i \cdot \nabla\right) \vec e_z \times \vec v_{i
\bot},
\]
and the stress tensor drift term  $\vec v_{\pi i}=\vec e_z \times
\nabp\cdot \pi_i/(m_i n_i \Omega_i)$ from the other side.

It is seen that, as long as the magnetic field is homogeneous, we
have
\be \nabla\cdot (n_i\vec v_{*i})\equiv 0, \label{e5} \ee
 describing
a well known fundamental property. This is due to the fact that the
diamagnetic drift is a fluid effect and not a particle drift, and
therefore it can not contribute to the flux in the continuity
equation. It appears due to the gyration of ion particles in the
presence of density gradient, and without any macroscopic motion of
the ion guiding center.  However, if an inappropriate linearization
is done, it gives rise to the terms $(\vec k \cdot \vec v_{*i})
n_{i1}$ in the continuity equation, that provides a false  source of
the current-driven instability that may be seen in the literature.

The mentioned cancelation of terms is due to the convective
derivative part
\be
 (\vec v_i\cdot\nabla)\vec e_z\times \vec v_{i\bot}
 \label{e5a}
\ee
in the polarization drift $\vec v_{pi}$. The procedure is described
in detail in \citet{v3,weil}. Within the approximation of small
gradients of the equilibrium quantities, the last $\vec v_{i\bot}$
in  the convective derivative (\ref{e5a}) contains only the leading
order perturbed drifts  from Eq.~(\ref{e3}). On the other hand, the
first $\vec v_i$ in (\ref{e5a})  can only be  the equilibrium ion
diamagnetic drift. This is then to be used in the term
$\nabla_\bot\cdot(n_i \vec v_{pi})$ in the continuity equation.

The stress tensor drift term  yields
\be \nabp\cdot(n \vec v_\pi) =-\rho_i^2 \nabp n_{i0} \cdot \nabp^2
\vec v_{i\bot} - n_{i0} \rho_i^2 \nabp^2 \nabp \cdot \vec v_{i\bot}.
\label{e6} \ee
Here, $ \quad \rho_i=\vti/\Omega_i$.  Within the second-order small
terms approximation, the first term on the right-hand side in this
expression {\em cancels out exactly} with the  contribution of the
above discussed convective derivative in the polarization drift
$\nabp\cdot[n  (\vec v_i\cdot\nabla)\vec e_z\times \vec v_{i\bot}]$.
The cancelation of terms is exact and valid for any plasma.
Moreover, the results obtained from this formal fluid theory  can
easily be obtained  by using the kinetic theory as well [cf.\
\citet{v3,weil}]. However, if the derivation is performed
incorrectly, by simply ignoring the stress tensor contribution, then
the resulting equations contain extra terms originating from the
convective derivative in the ion polarization drift. For
perturbations of the form $\sim \exp(- i \omega t + i k_y y + i k_z
z)$  this implies the presence of terms like  $\vec k \cdot \vec
v_{*i}$,  which in reality cancel out exactly.  A drastic example
where a false 'new instability' is obtained due to these terms may
be seen in \citet{mm}.

\subsection{Dissipative instability }

In the limit of perturbations with a parallel (to the magnetic field
vector) phase velocity that is considerably larger than the sound
speed $\omega/k_z\gg c_s= (\kappa T_e/m_i)^{1/2}$, the  ion
continuity equation reads \citep{v3}:
\[
\ddt \left(\frac{n_{i1}}{n_{i0}}\right) + \frac{1}{B_0} \vec e_z
\times \nabp \phi_1\cdot \frac{\nabp n_{i0}}{n_{i0}}
\]
\[
- \nu_i \rho_i^2 \nabp^2 \left(\frac{e \phi_1}{\kappa T_i} +
\frac{n_{i1}}{n_{i0}}\right) -  \rho_i^2 \ddt \nabp^2  \left(\frac{e
\phi_1}{\kappa T_i} + \frac{n_{i1}}{n_{i0}}\right)
\]
\be
 +
\frac{\rho_i^2}{\Omega_i} \ddt \nabp^4 \left(\frac{\phi_1}{B_0} +
\frac{v_{{\sss T}i}^2}{\Omega_i}\frac{n_{i1}}{n_{i0}}\right)=0.
\label{e8} \ee
Here, the discussion from the previous section is included
self-consistently.  For the assumed shape of perturbations,
Eq.~(\ref{e8}) yields:
\be
\frac{n_{i1}}{n_{i0}}= - \frac{\omega_{*i}+ \omega \rho_i^2 k_y^2
(1+ \rho_i^2 k_y^2) + i \nu_i \rho_i^2 k_y^2}{\omega[1+ \rho_i^2
k_y^2(1+ \rho_i^2 k_y^2)] + i \nu_i \rho_i^2 k_y^2}\,
\frac{e\phi_1}{\kappa T_i}, \label{e9} \ee
\[
 \omega_{*i}= k_y
\frac{\vti^2}{\Omega_i} \frac{n_{i0}'}{n_{i0}}, \quad \nabp n_{i0}=
- \vec e_x n_{i0}'=-\vec e_x  d n_{i0}/dx.
\]
The electron perpendicular and parallel velocities are obtained from
Eq.~(\ref{e2}):
 \[
  v_{e\bot}= \frac{1}{1+
\nu_{en}^2/\Omega_e^2}\left[\frac{1}{B_0}\vec e_z\times \nabla_\bot
\phi +\frac{\nu_{en}}{\Omega_e} \frac{\nabla_\bot \phi}{B_0} \right.
\]
\be \left. - \frac{v_{\sss{T} e}^2\nu_{en}}{\Omega_e^2}
\frac{\nabla_\bot n_e}{n_e}  -  \frac{v_{\sss{T} e}^2}{\Omega_e}
\vec e_z\times \frac{\nabla_\bot n_e}{n_e}\right].  \label{e10} \ee
\be
v_{ez1}=\frac{i k_z \vte^2}{\nu_{en}}  \left(\frac{e \phi_1}{\kappa
T_e} - \frac{n_{e1}}{n_{e0}} \right). \label{e11} \ee
From the electron continuity we then obtain
\be
\frac{n_{e1}}{n_{e0}}= \frac{\omega_{*e} + i D_p + i D_z}{ \omega  +
i D_p + i D_z } \frac{e \phi_1}{\kappa T_e}. \label{e12} \ee
Here $ D_p= \nu_e k_y^2 \rho_e^2$, $ D_z= k_z^2 \vte^2/\nu_e$,
$\rho_e=\vte/\Omega_e$, and
\[
 \omega_{*e}=-k_y \frac{\vte^2}{\Omega_e} \frac{n_{e0}'}{n_{e0}}.
 \]
The term $D_p$ describes the usually neglected effects of electron
collisions in the perpendicular direction. In most cases, in view of
the small inertia,  these electron collision effects are included
from the electron parallel momentum. This is justified provided that
\citep{v7}
\be k_z^2 \Omega_e^2/(k_y^2 \nu_e^2) \gg 1. \label{e13}
 \ee
In solar plasmas, for perturbations with an almost arbitrarily small
parallel wave number this condition may not always be satisfied.
This means that collisions must be taken into account in the
perpendicular dynamics.

In the absence of an equilibrium electric field and for
quasi-neutral perturbations, the dispersion equation of the drift
mode in collisional solar plasma is:
\[
 \frac{T_e}{T_i}\, \frac{\omega_{*i} + \omega \rho_i^2 k_y^2 (1+ \rho_i^2 k_y^2 ) +
 i \nu_i \rho_i^2 k_y^2}{\omega[ 1+ \rho_i^2 k_y^2 (1+ \rho_i^2 k_y^2)] +
 i \nu_i \rho_i^2 k_y^2}
\]
\be
 + \frac{\omega_{*e} + i (D_p + D_z)}{\omega + i (D_p + D_z)}=0. \label{e14}
\ee
In various limits, Eq.~(\ref{e14}) yields different dispersion
equations for the drift wave known from the literature.

The most simple case is for  collision-less plasma, when the
right-hand side in Eq.~(\ref{e14}) reduces to 1. Assuming, in
addition, that  $1\gg \rho_i^2 k_y^2$, we have
\be
\omega_r=\omega_{*e}/(1+ \rho_s^2 k_y^2), \quad \rho_s=c_s/\Omega_i,
\quad c_s^2=\kappa T_e/m_i. \label{dw} \ee
In the case of cold ions and for $D_z\gg D_p$, we obtain from
Eq.~(\ref{e14}) the same oscillation frequency as in the previous
example. But the growth rate due to electron collisions is
\citep{v6,weil}
 \be
\gamma=\frac{\nu_e  \omega_r^2 \rho_s^2 k_y^2}{k_z^2 \vte^2}.
\label{e15} \ee
As mentioned earlier, this describes the drift mode growing in the
presence of electron collisions. The full Eq.~(\ref{e14}) can be
solved numerically, taking relevant parameters for the solar
atmosphere at various altitudes \citep{v3}. The result represents an
interplay between the  ion collisions that damp the mode, and the
electron collisions that produce the growth rate approximately given
by Eq.~(\ref{e15}). This may have a great importance in terms of
plasma heating because the wave is always present in inhomogeneous
plasmas, and it grows on the account of collisions of light species,
while in the same time the wave energy is continuously absorbed by
the collisions of heavy species. The process continues till it runs
out of energy. In other words, it continues as long as the source
(i.e., the  density gradient)  is present.

\section{Kinetic instability of drift wave}

In most of the corona the collisions are not expected to play an
important role and consequently the drift dissipative instability
may not be of much importance.  In addition, regarding the problem
of coronal heating as a background and motivation  for any wave
analysis, in the case of so limited collisions an efficient
mechanism for transfer of energy from the wave to plasma is missing.
However, this is not so within  the frame of the drift wave kinetic
theory. In that case, the process develops as follows: the
interaction of the wave and electrons is destabilizing and the mode
grows due to Cherenkov-type interaction (in the presence of the
density gradient), while in the same time its energy is absorbed by
ions due to Landau damping. This may be seen from \citet{weil,ichi}
where, under the conditions
 \be
k_z \vti \ll\omega \ll k_z \vte, \quad \omega \ll \Omega_i, \quad
\!\!\frac{|k_y|}{|k_z|}\frac{\rho_i}{L_n}\left(\frac{T_e}{T_i}\right)^{1/2}\!\!
\gg 1, \label{cc} \ee
the wave properties  are described by the frequency
 \be
\omega_r=-\frac{\omega_{*i} \Lambda_0(b_i)}{1- \Lambda_0(b_i) +
T_i/T_e + k_y^2 \lambda_{di}^2},\label{k1} \ee
and the corresponding growth rate
\[
\gamma \simeq -\left(\frac{\pi}{2}\right)^{1/2} \frac{\omega_r^2}
{|\omega_{*i}| \Lambda_0(b_i)}\left[\frac{T_i}{T_e} \frac{\omega_r -
\omega_{*e}}{|k_z|\vte} \exp[-\omega_r^2/(k_z^2 \vte^2)] \right.
\]
\be
\left.
 +
\frac{\omega_r - \omega_{*i}}{|k_z|\vti} \exp[-\omega_r^2/(k_z^2
\vti^2)]\right]. \label{k2} \ee
Here,
\[
 \Lambda_0(b_i)=I_0(b_i) \exp(-b_i), \quad b_i=k_y^2 \rho_i^2,  \quad \lambda_{di}=\vti/\omega_{pi},
\]
and  $I_0$ is the modified Bessel function of the first kind and of
the order 0. The first condition in (\ref{cc}) refers to the use of
approximative expressions for the plasma dispersion function, while
the second condition implies a strongly magnetized plasma, and the
third implies that the acoustic part in the dispersion equation can
be omitted. Eqs.~(\ref{k1}) and (\ref{k2}), as well as
Eq.~(\ref{e14}),  are obtained using a local approximation. This
implies that the characteristic length for the change of the mode
amplitude  in the direction of the density gradient, is much larger
than the perpendicular wavelength, i.e., $(d/dx)^{-1}\simeq L_n\gg
\lambda_y$. Otherwise, an eigen-mode analysis is to be used, see for
more details in  \citet{vr04} and  \citet{veig}. Note that  in the
appropriate limits ($k_y^2 \lambda_{di}^2\ll 1$, $b_i\ll 1$)
Eqs.~(\ref{dw}) and (\ref{k1}) coincide. In Eq.~(\ref{k2}), the last
term in the square bracket is positive. It describes the damping on
ions and introduces a threshold in the mode instability that is of
importance only in plasmas with hot ions, yet negligible as long as
$\omega_r/k_z\gg \vti$ (hence  the reason for interest in small
$k_z$). The necessary condition for the instability follows from the
first term where we must have $\omega_r<\omega_{*e}$,  that is as a
rule easily satisfied. Equations~(\ref{dw}) and (\ref{k1}) reveal
the presence of the energy source already in the real part of the
frequency $\omega_r\propto \nabla_\bot n_0$, while details of its
growth due to the same source are described by Eqs.~(\ref{e15}) and
(\ref{k2}).

As an example,  in Fig.~3 the growth rate (\ref{k2}) is calculated
for an electron-proton plasma in terms of the parallel wavelength
$\lambda_z$ by taking $B_0=10^{-2}\;$T, $n_0=10^{15}\;$m$^{-3}$,
$L_n= [(dn_0/dx)/n_0]^{-1}=s\cdot 100\;$m, and for $\lambda_y=0.1$,
$0.3$, $0.5$, and $1\;$m. Because $k_y\gg k_z$, the corresponding
wave frequencies are practically constant and have approximate
values $186$, $210$, $254$, and $307\;$Hz, respectively (this
assuming $s=1$, but see below). In all these cases, we  have very
strongly growing drift modes. Note that in all four cases
$\omega_r<\omega_{*e}$, and the growth rate changes sign because of
the varying (with $\lambda_z$) contribution of the ion part in
Eq.~(\ref{k2}). Observe the extremely fast growth for short
perpendicular wavelengths $\lambda_y$. The growth rate increases
with $\lambda_z$ and may easily become much larger than $\omega_r$.
Yet, strictly speaking, in that case the assumption of smallness of
the imaginary part with respect to the real part, used in the
expansion of the plasma dispersion function, is violated, and the
problem must be treated numerically. In the examples presented in
Fig.~3, the conditions used in obtaining Eq.~(\ref{k2}) are formally
well satisfied, i.e.,  $\omega_r\ll \Omega_i$ and in the given range
of $k_z$, we have  $k_z \vti < \omega_r< k_z \vte$, and also the
sound branch is always far below the drift branch. For the given
parameters, the Debye length $\lambda_d$ is around $1\;$mm only, and
even shorter values of  $\lambda_y$ are permissible. The local
approximation is also well satisfied and the mode growth is expected
to take place throughout the density gradient. For the given
density, the plasma-$\beta$  is around  $0.6 m_e/m_i$.

\begin{figure}
\includegraphics[height=7cm, bb=15 15 293 220, clip=,width=.95\columnwidth]{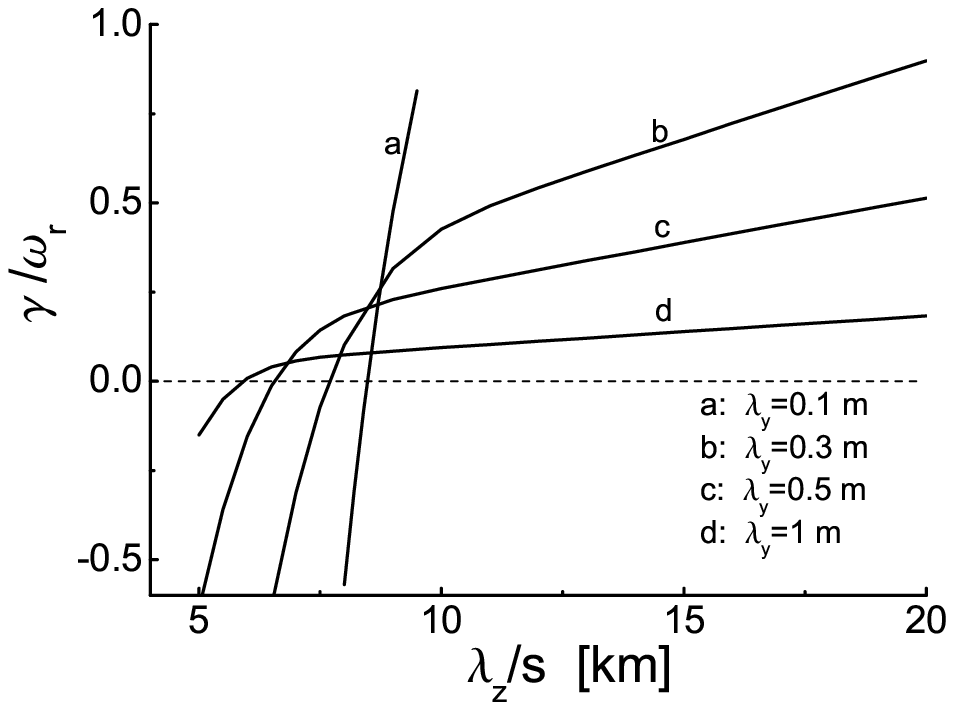}
 \caption{The growth rate (\ref{k2}) normalized to
the wave frequency $\omega_{r}$  in terms of the parallel
wavelength, for several perpendicular wavelengths $\lambda_y$, and
for the density scale-length $L_n=s\cdot 100\;$m, $s\in (0.1,
10^3)$. }\label{fig3}
 \end{figure}

A simple  way to demonstrate that a similar mode behavior  can take
place at various inhomogeneity scale-lengths $L_n$ (in other words
at various places in the corona), is to keep the ratio
$\lambda_z/L_n$ fixed. Thus, with the same parameters as above we
set $L_{n1} =  s\cdot L_n$, $\lambda_{z1} =  s\cdot\lambda_z$, where
$s$ takes values, e.g., between 0.1 and $10^3$. It can easily be
shown that the graphs from Fig.~3 remain {\em exactly the same}. In
other words, the ratio $\gamma/\omega_r$ remains unchanged, although
both $\gamma$ and $\omega_r$ are  shifted towards lower values. For
example, for $s=10^3$, i.e., $L_{n1}=10^5\;$m,
$\lambda_{z1}=10^7\;$m, and taking $\lambda_y=0.5\;$m, we have
$\gamma/\omega_r=0.26$. This is the same as the value of the line
labeled $c$ in Fig.~3, for $\lambda_z=10\;$km (and correspondingly
$L_n=100\;$m). Yet, now $\gamma=0.07\;$Hz, and $\omega_r=0.25\;$Hz.
Such a variation of $s$ may be used to describe the natural change
of the radial density gradient when we move  {\em along} a magnetic
flux tube, that may appear due to the diverging tube geometry. So
the mode growth occurs everywhere along the given flux tube. Hence,
as long as the conditions (\ref{cc}) used in the derivations are
satisfied, the results in Fig.~3 are valid for any $s$ in the range
$0.1 - 10^3$.

\begin{figure}
\includegraphics[height=7cm, bb=15 14 293 220, clip=,width=.95\columnwidth]{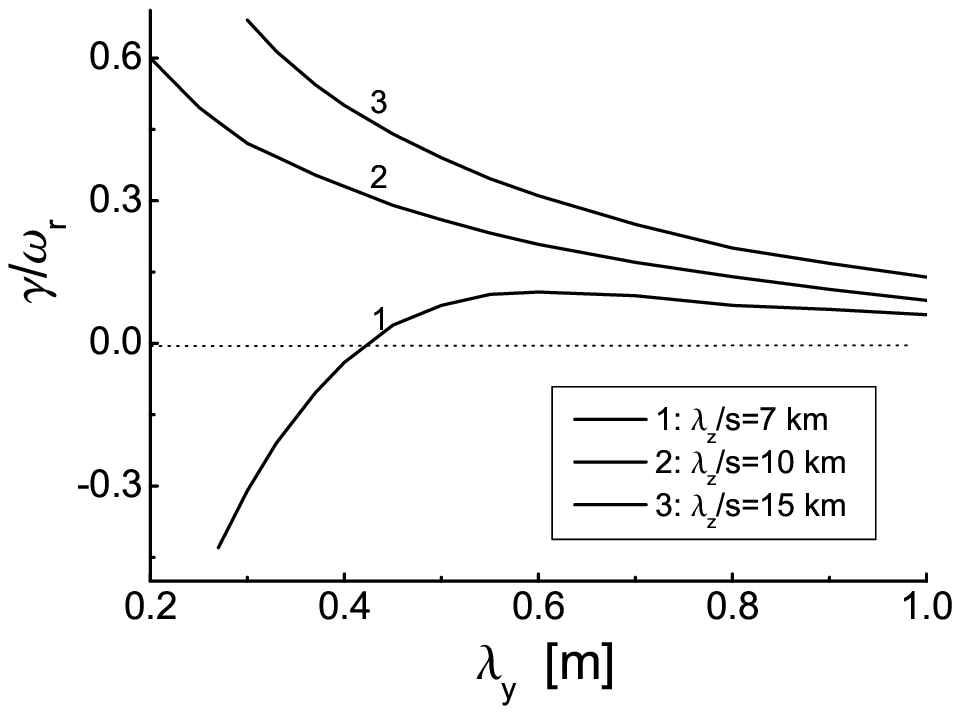}
 \caption{The growth rate (\ref{k2}) normalized to
$\omega_r$ in terms of the perpendicular wavelength, for several
values of $\lambda_z$. }\label{fig4}
\end{figure}

Next, we  check the mode behavior with respect to the perpendicular
wavelength  $\lambda_y$.  We  fix  $L_n=s\cdot 100\;$m, keep the
other parameters the same as above, and calculate the growth rate
for $\lambda_z/s=7$, $10$, and $15\;$km.  The result  is presented
in Fig.~4. The graphs and  the damping in the short $\lambda_y$
range (for $\lambda_z=7\;$m) are in agreement with Fig.~3.

\begin{figure}
\includegraphics[height=7cm, bb=15 15 293 220, clip=,width=.95\columnwidth]{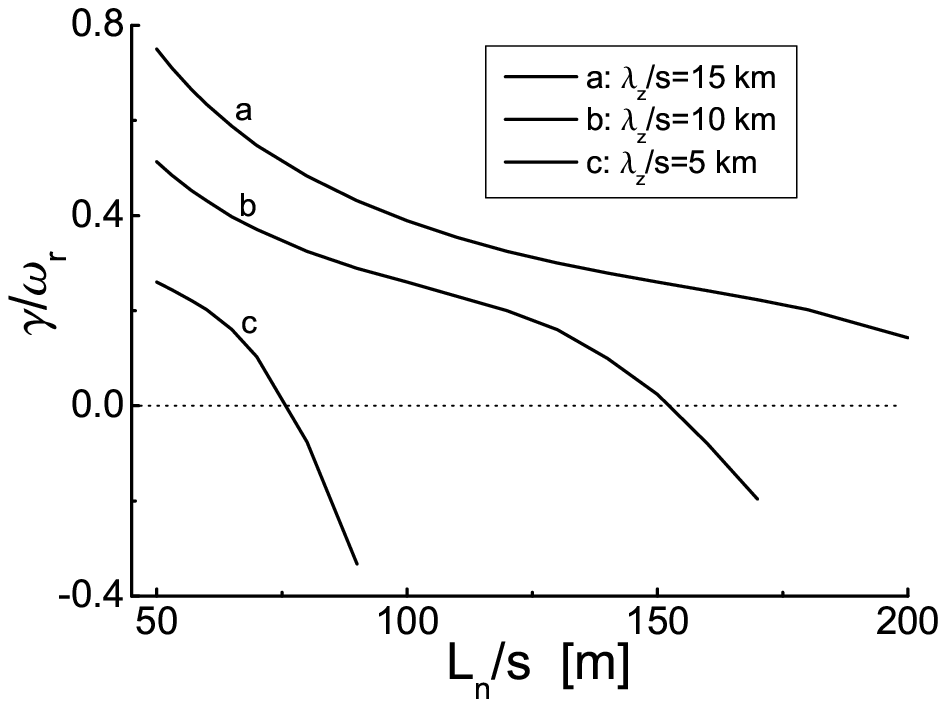}
 \caption{The growth rate (\ref{k2}) normalized to
$\omega_r$ in terms of the density scale length $L_n$  for several
values of $\lambda_z$, $s\in (0.1, 10^3)$. }\label{fig5}
\end{figure}

In Fig.~5 we present the mode growth rate in terms of $L_n$, for the
same plasma parameters as above.   The graphs remain unchanged for
any $s\in (0.1, 10^3)$. The frequency is  $507\;$Hz (at
$L_n=50\;$m), and $254\;$Hz (at $L_n=170\;$m), for  $s=1$. In the
case $s=10^3$, the frequency is $0.5\;$Hz (where now $L_n=50\;$km),
and $0.15\;$Hz (at $L_n=170\;$km).

A similar quantitative analysis may be preformed for densities
several orders of magnitude below the used value, and for at least
one order of magnitude higher density as well, resulting in the same
wave behavior, with only some shift in the wave frequencies and
wavelengths. This is seen from Fig.~6, where the growth rate is
plotted in terms of $\lambda_y$ and $s\cdot L_n$ by taking
$n_0=10^{13}\;$m$^{-3}$,  $B_0=10^{-3}\;$T, and $\lambda_z=s\cdot
2\cdot 10^4\;$m.  The smaller values of $n_0$ and $B_0$ imply the
possible application of the model to  higher altitudes in the
corona. It is seen that the perpendicular wavelength is just shifted
up for one order of magnitude, and the growth rate decreases with
$\lambda_y$ and $L_n$ just like in Figs.~4 and 5.

On the other hand, for densities that are two or more orders of
magnitude larger than the earlier used value $10^{15}\;$m$^{-3}$,
and thus for a larger plasma-$\beta$, a coupling with
electromagnetic Alfv\'en-type perturbations may take place
\citep{has2}. As  shown  elsewhere \citep{v3}, the drift mode
behavior will remain similar even in that case. The  frequency
becomes slightly reduced and  a part of the wave energy is  spent on
the coupling with this additional Alfv\'en mode that appears to be
always damped by collisions. This coupling is described by the
following dispersion equation   \citep{weil,v3}
\[
\omega^3 - \omega^2 (\omega_{*e}+\omega_{*i}) + \omega
[\omega_{*e}\omega_{*i} - k_z^2 c_a^2/(1+ k_y^2 \rho_i^2)
\]
\be - k_y^2 k_z^2 c_a^2 (\rho_i^2 + \rho_s^2)] + k_z^2 c_a^2
\omega_{*e}/(1+ k_y^2 \rho_i^2) =0.\label{coup} \ee
Here, $c_a^2=B_0^2/(\mu_0 n_0 m_i)$. The nature of coupling is best
seen for negligible ion thermal effects when we have $
(\omega-\omega_{*e}) (\omega^2 - k_z^2 c_a^2)- \omega  k_z^2 c_a^2
k_y^2 \rho_s^2=0$. For sufficiently small  $k_y^2\rho_s^2$,  the two
modes propagate practically almost independently.

\begin{figure}
\includegraphics[height=7cm, bb=40 20 305 240, clip=,width=.95\columnwidth]{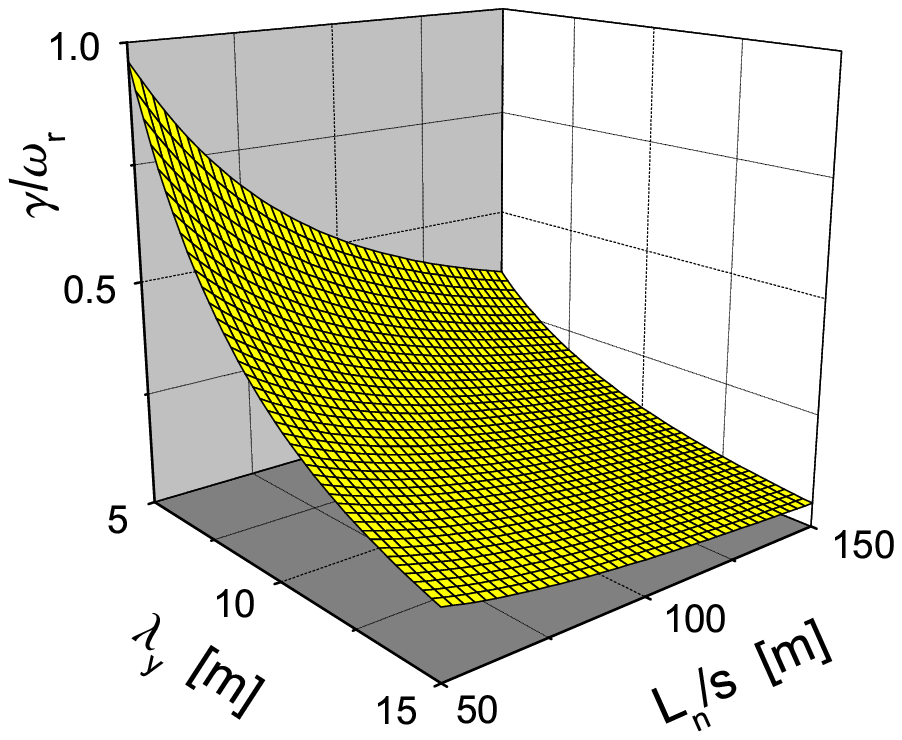}
\caption{The growth rate (\ref{k2}) normalized to the
wave frequency $\omega_{r}$  in terms of the perpendicular
wavelength $\lambda_y$ and  the density scale-length $L_n$, for
$\lambda_z= s \cdot 2\cdot 10^4\;$m,  $s\in (0.1, 10^3)$.
}\label{fig6}
\end{figure}

In view of all these results,  we stress again  an important
difference of the drift wave as compared to other plasma modes. The
mode frequency given by Eq.~(\ref{k1}), already by its form {\em
implies} the presence of an energy source stored in the density
gradient. This free energy is then responsible for its growth,
either due to kinetic or fluid effects, and the details of the
growth are described by  Eqs.~(\ref{e15}) and (\ref{k2}). Compare
this with, e.g., an Alfv\'{e}n  $\omega=k c_a$, or a sound mode
$\omega=k c_s$, where the given dispersion equations only allow for
the {\em possibility } for the plasma to support these modes, on the
condition that an {\em additional} energy source is provided.

As for the experimental verification  of such a very strongly
growing drift-wave instability, one (out of many) may be found e.g.,
in \citet{br}  for a similar almost collision-less {\em hydrogen}
plasma with $T_e=14\;$eV, $T_i=2\;$eV, $B_0\leq 0.35\;$T, and the
density $n_0=10^{16}\;$m$^{-3}$. Note that in the experiment the
maximum observed linear growth-rate was very high: $\omega_i\simeq
\omega_r=7\cdot 10^5\;$Hz.

\subsection{Electron acceleration by parallel wave-electric field}

The electrostatic drift mode presented in Figs.~3-6 implies a
time-varying electric field, whose parallel component $|\nabla_z
\phi|/E_d=|k_z \phi|/E_d$, normalized to the Dreicer runaway
electric field, for  $\lambda_z=10\;$km and  $\phi=60, 80\;$V/m, is
$3.1$ and  $4.1$, respectively.  Here, we use $\lambda_y=0.5\;$m,
$B_0=10^{-2}\;$T, $n_0=10^{15}\;$m$^{-3}$, $L_n=
[(dn_0/dx)/n_0]^{-1}= 100\;$m, and the reason for the given values
of $\phi$ will be given in the forthcoming sections. The Dreicer
electric field is \citep{dr} $E_d=e L_{ei}/(4 \pi \varepsilon_0
\lambda_d^2)$. Here, $L_{ei}=\log(\lambda_d/b_0)$ is the Coulomb
logarithm, $\lambda_d=\lambda_{de} \lambda_{di}/(\lambda_{de}^2+
\lambda_{di}^2)^{1/2}$ is the plasma Debye radius,  and
$b_0=[e^2/[12 \pi \varepsilon_0 \kappa (T_e+ T_i)]$ is the impact
parameter for electron-ion collisions \cite{vx}. For the given
temperature of one million K, we have $L_{ei}=20.1$,
$\lambda_d=0.0015\;$m, and the Dreicer field is $0.012\;$V/m. Hence,
the parallel wave field  exceeds the Dreicer field so that  the bulk
plasma species (primarily electrons) can be accelerated/decelerated
by the wave in the parallel direction.  The acceleration is more
effective on particles that are already more energetic, resulting in
a distribution function considerably different from  a maxwellian.
This may be one  of the reasons behind the observed
kappa-distribution in the outer solar atmosphere and in the solar
wind.

For $s=10^3$ and for the two values of $\phi$ mentioned above, the
normalized electric field  is $0.003$ and $ 0.004$, respectively.
For such small normalized values, and also in view of the fact that
the electron mean-free-path is orders of magnitude below the
parallel wave-length,  the effect of the acceleration on the bulk
electrons is negligible.

\begin{figure}
\includegraphics[height=7cm, bb=15 15 292 222, clip=,width=.95\columnwidth]{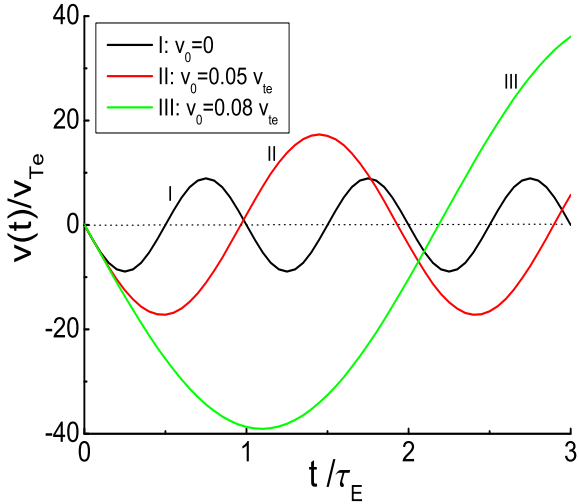}
 \caption{Velocity of electrons (\ref{dv}) accelerated
by the drift-wave electric field with $\phi=80$ V, for three
different starting velocities. }\label{fig7}
\end{figure}

The electron parallel velocity in such a time-varying wave-electric
field $E_0 \cos (k_z z - \omega_r t)$, is approximately given by
\cite{bit}
\[
v(t)= v_0-\frac{e E_z}{m_e(k_z v_0 - \omega_r)} \left\{ \sin[k_zz_0
+ (k_z v_0 - \omega_r) t] \right.
\]
\be \left.
- \sin(k_zz_0)\right\}. \label{dv} \ee
Here, $v_0$ and $z_0$ are the starting electron velocity and
position in the parallel direction, respectively. Clearly, the
acceleration of every separate particle is dependent on its
particular velocity $v_0$, the lowest being for those with $v_0=0$.
A strong acceleration will take place for resonant particles
satisfying $v_0=\omega_r/k_z$ [see also in  \citet{fle}]. For the
parameters used above $\lambda_z=10\;$km and $\lambda_y=0.5\;$m, we
have $\omega_r=254\;$Hz, and $\gamma/\omega_r=0.26$. Thus, the
resonant particles are those with  $v_0=404\;$km/s $\simeq 0.1
\vte$.

The acceleration of electrons with different starting velocities is
seen in Fig.~7, where the achieved velocity $v(t)$ (normalized to
$\vte$) is presented for three electron populations with
$v_0/\vte=0$, $0.05$, and $0.08$. The figure describes the different
effective oscillation periods $2 \pi/(k_z v_0- \omega_r)$ and the
amplitudes of the three electron populations within the same wave
period $\tau_{{\sss E}}=2 \pi/\omega_r=0.025$~s of the electric
field with $\phi=80$ V (which coincides with the electron population
with $v_0=0$).

The given  time variation implies electron fluxes in both directions
along the $z$-axis. For the line labeled III in Fig.~7, we have the
electron velocity $\Delta v\simeq 40 \vte$ corresponding to the
energy of $10^{-14}$ J, or around 69 keV. This oscillatory
acceleration/deceleration is expected to considerably contribute to
electron heating only if $\nu_e\gg \omega_r$. The process is also
accompanied by a radiation, with the energy \citep{pan} radiated by
an electron
\[
 \frac{\Delta \Sigma}{\Delta t}=\frac{1}{4 \pi
\varepsilon_0} \frac{2 e^2}{3 c^3} \frac{(\Delta v/\Delta t)^2 -
(\vec v\times \Delta \vec v/\Delta t)^2/c^2}{(1- v^2/c^2)^3}.
\]
Setting as an example $\Delta t=\tau_{{\sss E}}$, for this
particular  case we obtain  a very small  energy  $\Delta \Sigma=1.4
\cdot 10^{-35}\;$J radiated by an electron in one act of
deceleration.

Because the wave-field is, in fact, created by the plasma particles,
this acceleration/de\-ce\-le\-ra\-ti\-on should have a  feed-back
effect onto the wave itself. Clearly, it may affect  the wave
amplitude, especially for shorter wavelengths. In this case, the
escaping electrons imply a lower amount of them remaining to shield
the ion perturbations, and the perturbations should be increased.
Yet, this all could be captured consistently only by numerical
tools.

\subsection{Plasma  heating by the drift wave}

From Eq.~(\ref{k2}), it is seen that for a wave frequency below
$\omega_{*e}$, the dissipation rate can formally be written as
$\omega_i=|\gamma_{el}|-|\gamma_{ion}|$, and two parallel mechanisms
of plasma heating are in action here.

The  term $|\gamma_{ion}|$ is responsible for the Landau dissipation
of the wave energy and, consequently, for the heating of the plasma.
So, as long as the density gradient is present, there is a
continuous precipitation of energy from the wave to the plasma. Note
that similar heating due to the Landau dissipation of the ion
acoustic mode \citep{da} predicts a stronger heating of ions
\citep{rev}, in agreement with observations.

On the other hand, the term $|\gamma_{el}|$ produces a growth of the
wave, and this implies another (stochastic) heating mechanism that
also involves single particle interaction with  the wave. This is a
process described and {\em experimentally verified} in
\citet{san2,san}. For drift wave perturbations of the form $\phi(x)
\cos(k_y y + k_z z - \omega t)$, $|k_y|\gg |k_z|$, one finds the ion
particle trajectory in the
wave field from the following set of equations: %
\[
d \chi/d \tau=\Upsilon, \quad \quad \chi=k_y x, \quad \Upsilon= k_y
y, \quad \tau=\Omega_i t
\]
\be
d^2 \Upsilon/d \tau^2=- \Upsilon + [m_i k^2 \phi/(e B_0^2)]
\sin(\Upsilon- \tau \omega/\Omega_i). \label{ht1} \ee
 The analysis
from \citet{san} reveals that stochastic heating takes place for a
sufficiently large wave amplitude, more precisely for
\be
a= k_y^2 \rho_i^2 \,\frac{e \phi}{\kappa T_i} \geq 1. \label{h2} \ee
The maximum achieved bulk ion velocity is shown to be proportional
to the wave amplitude and is given by
 \be
  v_{max}\simeq [k_y^2 \rho_i^2 e
\phi/(\kappa T_i) + 1.9]\Omega_i/k_y. \label{vm} \ee
Ideally, this requires \be
 |\gamma_{el}|\geq |\gamma_{ion}|,\label{h1}
\ee
so that the wave amplitude may grow and at some point both heating
mechanisms may  take place, simultaneously. The condition (\ref{h1})
can easily be satisfied in view of the almost unlimited  range of
the parallel wave number $k_z$, so that the ion Landau damping can
be made  small, i.e., $|\omega/k_z|\gg \vti$.

In the stochastic heating  due to the drift wave,  the ions move in
the perpendicular direction to large distances and feel the
time-varying field of the wave due to the polarization drift $ \vec
v_p=(\partial \vec E/\partial t)/(\Omega_i B_0)$ [the sixth term in
Eq.~(\ref{e3})], and as a result their motion becomes stochastic. In
other words, the polarization drift of the ions becomes comparable
to the $\vec E\times \vec B$ drift, and the displacement due to the
polarization drift is comparable to the wavelength. The single-ion
motion becomes  chaotic because of trapping (in the  wave potential
well) and de-trapping due to the magnetic field. The polarization
drift is in the direction of the wave number vector, which
emphasizes the crucial {\em electrostatic nature} of the wave in the
given process of heating.   Also important  to stress is that in
this scenario the  stochastic heating is  highly anisotropic, and it
takes place mainly in the direction normal to the magnetic field
$B_0$ (both the $x-$ and $y$-direction velocities are stochastic).
In the same time, in view of the mass difference and the physical
picture given above, this heating scenario predominantly acts on
ions. The heating is maximal in the areas of strong density
gradients (the areas of maximal drift wave activity), and also
proportional to the strength of the magnetic field (the stronger the
field the more localized heating).  All these facts have been
confirmed experimentally and, in the solar case, satisfy the
observational constraints for the coronal heating mechanism
discussed in the introduction.

 As a matter of fact, in application to coronal plasma, the
indication or proof that the described heating really takes place
would be: i)~an ion temperature anisotropy $T_{i\bot}\gg T_{i z}$,
ii)~a possibly higher ion temperature in comparison to electrons,
and iii)~a better heating of heavier ions.  Observations show that
i)  may be taken rather as a rule than as an exception [see in
\citet{li},  \citet{cus}, \citet{cr3}], i.e., the perpendicular
stochastic heating is more dominant compared to the parallel
heating. This may also be expected from Fig.~3, where the Landau
resonance $\omega/k_z\sim \vti$ takes place for short wavelengths.
For example, in the case of line $c$ there, having $\vti\simeq
91\;$km/s, the ion resonance takes place at $\lambda_z\simeq
2.2\;$km, i.e., in the domain where the wave is strongly damped and
will not appear at all. There are also numerous indications that
confirm the features ii) and iii). As an example, we refer to graphs
from \citet{hans}, where $T_e<T_H<T_{He}$ throughout  the corona and
the solar wind.  Similar results may also be seen  in \citet{cus}
and \citet{cr3}, and in references cited therein.

A stronger heating of heavy ions can be understood from
Eq.~(\ref{vm}) and after expressing the effective temperature in
terms of the ion mass $T_{eff}(m_i)=m_iv^2_{max}/(3\kappa)$. From
the derivative $dT_{eff}(m_i)/dm_i>0$, we find that the heating
increases with the ion mass if:
\be
k^4_y\rho^4_i\left( \frac{e\phi}{\kappa T_i}\right)^2 >
1.9.\label{cond} \ee
For $\lambda_y=0.5\;$m, $\lambda_z=10\;$km, $L_n=100\;$m, we obtain
from Eq.~(\ref{k2}) $\omega_i/\omega_r=0.26$ and $\omega_r=254\;$Hz.
Note that in this case $a\simeq 1$ and the stochastic heating is in
action. Assuming small starting perturbations $e \phi/(\kappa
T_i)=0.01$, i.e., $\phi=0.86\;$V, the value $\phi=60\;$V is achieved
within $\tau_g= 0.06\;$s. Those were  the reasons for the amplitudes
of $\phi$ used in the preceding section.

For $\phi=60\;$V we then have $T_i(\lambda_y, \mu)=0.881 + 0.057
\mu/\lambda_y^2
   + 1.78 \lambda_y^2/\mu$ (normalized to the starting temperature $T_i=10^6$ K). In Fig.~8, we plot the
obtained temperature in terms of the ion mass (normalized to the
proton mass), and the perpendicular  wavelength $\lambda_y$. It is
seen that for short  $\lambda_y$, and therefore for fast growing
modes, {\em the heating is always larger for heavier ions}.

\begin{figure}
\includegraphics[height=7cm, bb=45 16 293 230,clip=,width=.95\columnwidth]{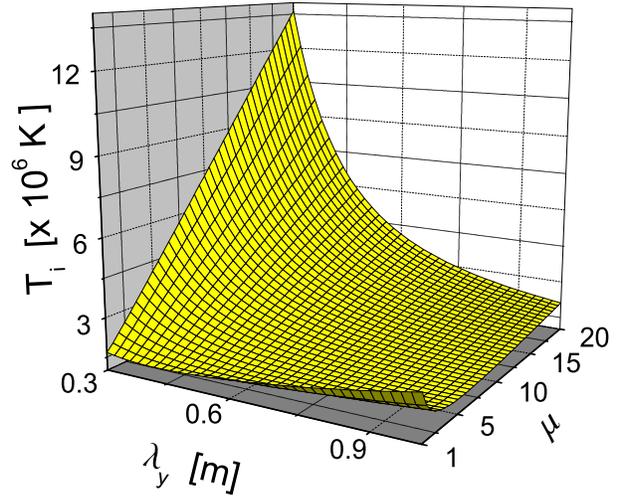}
\caption{The stochastically increased ion temperature
$T_{eff}=m_i v_{max}^2/(3 \kappa)$ (in millions K) in terms of the
perpendicular wave-length $\lambda_y$ and the ion mass.
}\label{fig8}
\end{figure}
Using the same starting set of parameter values as above, by using
Eqs.~(\ref{h2}) and (\ref{vm}), in Table~1 we calculated the
effective increase of the ion temperature for hydrogen and helium
for several wave amplitudes and perpendicular wavelengths. The
results confirm and quantify the conclusions drawn above. The
stronger heating for helium (values in brackets) in the short
wavelength range is because the condition (\ref{cond}) is easily
satisfied. Obviously, the proposed mechanism is more than efficient
enough to heat the ions in the solar corona to the observed
temperatures.

Assuming  again small perturbations $e\phi/(\kappa T_i)=0.01$ for
$\lambda=0.5\;$m, $L_n=100\;$m we have the above given values for
$\omega_i, \omega_r$, and the value $\phi=60\;$V  achieved within
$\tau_g= 0.06\;$s. The maximum energy released per unit volume is
$\Sigma_{max}=n_0 m_i v_{max}^2/2=0.04\;$J/m$^3$. The energy release
rate $\Gamma_{max}=\Sigma_{max}/\tau_g \simeq 0.7\;$J/(m$^3$\,s)
amounts to 4 orders of magnitude above the necessary value.

However, for $L_n=100\;$km (i.e., setting $s=1000$) we obtain
$\omega_i=0.07\;$Hz, $\omega_r=0.25\;$Hz, $\tau_g= 64\;$s, and
consequently $\Gamma_{max}=6.3\cdot 10^{-4}\;$J/(m$^3$\,s), that is
around 10 times the value presently accepted as necessary. Hence,
the heating rate in large magnetic loops comes close to the required
values. Similar estimates may be made for still larger $L_n$, yet
the conditions under which the previous expressions are derived
become violated and a numerical approach is required in this case.

For the  typical achieved effective temperatures  from Table~1,  we
have  a heating rate of ions of the order of $10^7-10^8\;$K/s, which
is similar to the heating rate obtained  in the experiments
\cite{san,san2}. Observe also that the magnitude of the electric
field which we are dealing with, is of the same order as in the
experiments.

 However, some effects that are not included here may
reduce the energy yield, especially at shorter spatial scales. They
read as follows. In reality, the nonlinearity \cite{lee} and
collisions lead to the radial flattening of the density profile in
the region occupied by the wave, resulting in the saturation of the
growth. The flattening in some  region $\triangle r$ occupied by the
wave, around a point $r_0$, leads  to the  saturation of the
instability in this particular region. However, the process is
accompanied by a simultaneous  steepening of the density profile
outside the region $r_0 \pm \triangle r/2$. These newly created (and
also even steeper) density gradients will support the excitation of
new modes now at different positions in radial direction. So here we
have a sort of 'double cascade': the regions affected by the
excitation of waves (and the heating) are shifting  radially in both
positive and negative directions. In other words, the starting
instability (and heating) initiated  around  the position $r_0$ (and
in the same time extraordinarily elongated in the axial direction
because $k_z/k_y\ll 1$) will have the tendency of spreading
radially.

The  energy diffusion (due to any reason) in the perpendicular
direction may reduce the local effects of the demonstrated strong
heating rate.  The earlier given Eq.~(\ref{coup}) describes the
coupling with the Alfv\'{e}n  wave, that is proportional to $k_z
k_y$. All these  effects (the nonlinearity, collisions, particle
acceleration by the electric field, diffusion, and coupling with the
Alfv\'{e}n wave) will more effectively act on short scales. All of
them, except the particle acceleration, will tend to reduce the mode
amplitude. The actual values for $\Gamma$ are thus expected to be
below $\Gamma_{max}$. Therefore, the apparently too large release of
energy at short scales, as formally obtained above, may in reality
be considerably reduced. Clearly, more accurate estimates and more
detailed description may be obtained only numerically.

\begin{table*}
 \centering
 \begin{minipage}{70mm}
  \caption{Plasma heating for hydrogen  and helium  (in brackets) ions for several values of the
 perpendicular wavelength and  the wave amplitude $\phi$. The
maximum stochastic velocity is given by  Eq.~(\ref{vm}) and the
effective temperature obtained by heating  is $ m v_{max}^2/(3
\kappa)$. }
  \begin{tabular}{@{}llr@{}}
  \hline
        &  $\phi=60$ [V]          &     \\
     \hline
  $ \lambda_y$ [m]     &  $v_{max} $ [m/s]       &  $ T_{eff}$ [K]  \\
        \hline
  0.3  & $2.12 \cdot 10^5$ ($1.50 \cdot 10^5$) & $1.82 \cdot 10^6$ ($3.50 \cdot 10^6$)      \\
  0.5  & $2.20 \cdot 10^5$ ($1.10 \cdot 10^5$) & $1.96 \cdot 10^6$ ($2.01 \cdot 10^6$)     \\
   0.8  & $2.79 \cdot 10^5$ ($1.78 \cdot 10^5$) & $3.14 \cdot 10^6$ ($1.78 \cdot 10^6$)     \\
\hline
 \hline
       & $ \phi=80 $   [V]          &     \\
    \hline
  $ \lambda_y$ [m]     &  $v_{max} $ [m/s]       &  $ T_{eff}$ [K]  \\
 0.3  & $2.54 \cdot 10^5$ ($1.89  \cdot 10^5$) & $2.61 \cdot 10^6$ ($5.78  \cdot 10^7$)      \\
   0.5  & $2.45 \cdot 10^5$ ($1.37\cdot 10^5$) & $2.43  \cdot 10^6$ ($3.02\cdot 10^6$)     \\
     0.8  & $2.95 \cdot 10^5$ ($1.21\cdot 10^5$) & $3.50  \cdot 10^6$ ($2.35\cdot 10^6$)     \\
 \hline\hline
     &   $ \phi=100$ [V]           &     \\
      \hline
  $\lambda_y$ [m] & $v_{max}$ [m/s] & $ T_{eff}$ [K]  \\
\hline
    0.3  & $2.96 \cdot 10^5$ ($2.30  \cdot 10^5$) & $3.54 \cdot 10^6$ ($8.62  \cdot 10^7$)      \\
    0.5  & $2.71 \cdot 10^5$ ($1.62\cdot 10^5$) & $2.95  \cdot 10^6$ ($4.23\cdot 10^6$)     \\
      0.8 & $3.10 \cdot 10^5$ ($1.36\cdot 10^5$) & $3.88  \cdot 10^6$ ($3\cdot 10^6$)     \\
        \hline
 \hline
\end{tabular}
\end{minipage}
\end{table*}

Some  additional features of the drift wave may also play a crucial
role. First, the mode becomes easily nonlinear, and, second, within
the nonlinear theory it allows a double cascade in $\vec k$-space,
i.e., the transport of the wave energy both towards large and short
wavelengths. Hence, due to nonlinear three-wave parametric
interaction, a mode growing for certain $\omega_0, \vec k_0$ will
tend to excite modes at rather different (both smaller and larger)
$\omega_j, \vec k_j$, and in the end those will always include the
modes that heat the system most effectively. For example, the shape
of Eq.~(\ref{vm}) reveals that $v_{max}\sim a_1 k_y + a_2/k_y$ and,
in terms of $k_y$, $v_{max}$ has a minimum at $k_c^2=1.9 \kappa
T_i/(\rho_i^2 e \phi)$. Hence, even if a mode with  wave number
$k_c$ does not produce a so large maximum bulk velocity $v_{max}$,
it will nonlinearly excite modes with $k_y$ far from  $k_c$ and
there will be a stronger heating with this  nonlinearly generated
mode.

\subsection{Heating of cool corona}

In sections 4.1 and 4.2, we have demonstrated the possibility for
sustainable  coronal temperatures of around  million K with the
discussed drift wave mechanism. Assume now a relatively cool
starting corona, with a temperature of only $10^4$ K (the value
taken high enough in order to neglect the presence of neutrals) and
let us calculate the time that is needed to achieve the temperature
of 1 million K, using the same mechanism.

We first chose  the least favorable value  $k_y=k_c$ calculated
above. This yields $a=1.9$ for any mode amplitude $\phi$, so we have
the stochastic heating in action. For these $a, k_c$, from
Eq.~(\ref{vm}) we obtain $v_{max}=3.8 \Omega_i \rho_i [e \phi/(1.9
\kappa T_i)]^{1/2}$. The effective temperature of 1 million K
implies that the mode amplitude has the value  $\phi=\phi_m=34\;$V.
This then yields $k_y=k_c=23.14\;$1/m and $\lambda_c=0.27\;$m.

Hence, choosing  $\lambda=\lambda_c$, and  $\lambda_z=50\;$km,
$L_n=100\;$m, and $n_0=10^{15}\;$m$^{-3}$,  from Eqs.~(\ref{k1}) and
(\ref{k2}) we find $\omega_r=18.3\;$Hz and $\gamma=0.75\;$Hz.
Assuming a small starting perturbation of the potential
$\phi=0.0086\;$V, i.e., $e \phi/(\kappa T_i)=0.01$, with the given
growth rate we find out that the mode amplitude $\phi_m$ (at which
the temperature achieves the required value of million K due to  the
stochastic heating), is reached at longest within $11\;$s. For any
other value of $\lambda_y$ we will have a shorter growing and
heating time.

\section{Kinetic current-driven drift wave instability}

In the presence of an additional energy source, in the form of an
electron current in the direction of the magnetic field vector, the
drift wave becomes even  more unstable as compared to the previous
collisionless kinetic instability. The mode frequency and the growth
rate, with some additional features and details \citep{hat}, are
then given by:
\[
\omega_r= \omega_{*e} \frac{\Lambda_0(b_i)}{1+ [1- \Lambda_0(b_i)]
\eta_\bot} \left[ 1 \right.
\]
\be \left. + \frac{1+ [1- \Lambda_0(b_i)] \eta_\bot + \Lambda_0(b_i)
\eta_z}{1+ [1- \Lambda_0(b_i)] \eta_\bot} \frac{k_z^2 \vtiz^2}{2
\omega_{*ep}^2} \right] \label{hat1} \ee
\[
\omega_i= \frac{\omega_r \pi^{1/2}}{1+ [1- \Lambda_0(b_i)]
\eta_\bot} \left[1+ \frac{\eta_z \Lambda_0 k_z^2
\vtiz^2/\omega_{*ep}^2}{2[ 1+ [1- \Lambda_0(b_i)] \eta_\bot]}
\right]
\]
\[
 \times \left\{ \frac{u_0}{\vte} + \frac{\omega_{*e} -
\omega_{*ep}}{k_z \vte} \right.  \left. - \Lambda_0(b_i)
\frac{\eta_z \omega_{*ep} + \omega_{*e}}{k_z \vtiz} \right.
\]
\be
\left. \times  \exp[- \omega_{*ep}^2/(k_z^2 \vtiz^2)] \right\}.
\label{hat2} \ee
Here, $u_0$ is the electron current, the starting ion temperatures
in the perpendicular and parallel direction are allowed to be
different, and we use the notation from \citet{hat} $
\eta_{\bot,z}=T_e/T_{i,\bot,z}$, $ \vtiz^2=\kappa T_{iz}/m_i$, and $
\omega_{*ep}= \omega_{*e} \Lambda_0(b_i)/[1+ (1-
\Lambda_0(b_i))\eta_\bot]$.

In solar corona environment, such electron currents may appear in
the processes of magnetic reconnection. An example with the same
phenomenon  for the terrestrial atmosphere is given in \citet{sato}.

We calculate the growth rate (\ref{hat2}) numerically by taking
$B_0=10^{-3}\;$T, $n_0=10^{13}\;$m$^{-3}$, $L_n=
[(dn_0/dx)/n_0]^{-1}= 1\;$km, $T_e=T_{i\bot}=10^6= 2 T_{iz}$, and
$\lambda_z=200\;$km. This  different set of parameters is chosen
only for a change, in view of the comments given earlier in Sec.~4
(see also Fig.~6).  The result is presented in Fig.~9, where the
growth rate $\omega_i(\lambda_y, u_0)$ is normalized to the wave
frequency $\omega_r(\lambda_y, u_0)$. We observe that for the given
set of parameters, the  current additionally increases the growth
rate. This is seen by comparing the limits $u_0=0$ and $u_0=10$, for
e.g. $\lambda_y=15\;$m,  where the growth rate is increased by about
a factor 20. However, the effect is of  less importance for shorter
wavelength (see the limit $\lambda_y=5\;$m). In  the given example,
the mode frequency is below $\omega_{*e}$, so that the wave is
growing anyhow, and this is due to the kinetic effects presented in
the previous sections.

The real importance of such an electron flow is completely different
in the limit $\omega\simeq \omega_{*e}$ when the instability sets in
provided that the electron current exceeds a certain threshold. This
may be checked for example by setting larger values for $B_0$ when
the  kinetic instability from the previous sections vanishes
(because for the given parameters $ \omega_{*e}$ is reduced so that
$\omega\geq \omega_{*e}$). Yet, the wave instability reappears for a
sufficiently large $u_0$, but it is now the current-driven one.
Exactly such a sort of behavior was {\em verified experimentally} in
\citet{hat}. Hence, for example we take the magnetic field for one
order of magnitude larger $B_0=10^{-2}\;$T, and the result is
presented in Fig.~10. The contour lines of $\omega_i(\lambda_y,
u_0)/\omega (\lambda_y, u_0)$ in Fig.~10 yield only the
current-driven drift wave for sufficiently large $\lambda_y$.

\begin{figure}
\includegraphics[height=7cm, bb=40 10 300 230,clip=,width=.95\columnwidth]{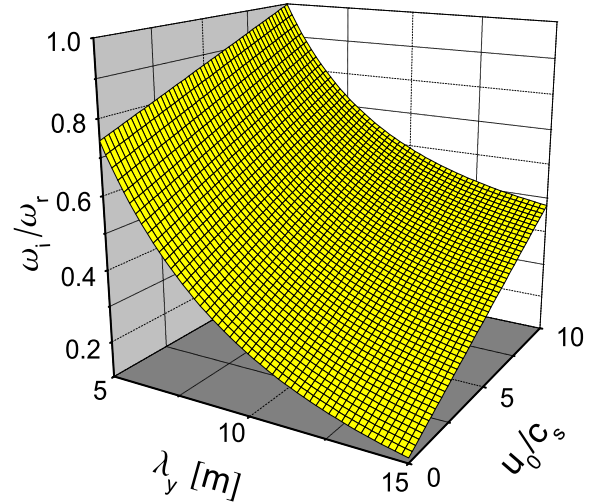}
 \caption{Growth rate $\omega_i(\lambda_y, u_0)/\omega
(\lambda_y, u_0)$ from Eqs.~(\ref{hat1}) and (\ref{hat2}) for the
kinetic + current-driven drift-wave instability  in the case
$B_0=0.001\;$T. }\label{fig9}
\end{figure}

\begin{figure}
\includegraphics[height=7cm, bb=10 10 300 230,clip=,width=.95\columnwidth]{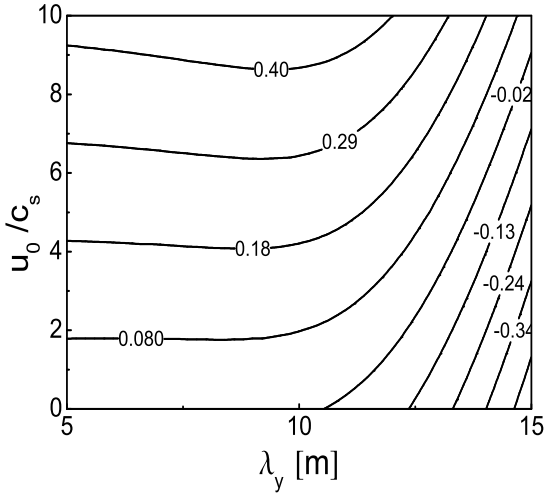}
 \caption{Contour lines of $\omega_i(\lambda_y,
u_0)/\omega (\lambda_y, u_0)$ from Eqs.~(\ref{hat1}) and
(\ref{hat2}) for $B_0=0.01\;$T, showing  the current-driven
drift-wave instability in the vicinity of threshold.}\label{fig10}
\end{figure}

\subsection{Plasma heating}

The heating of the plasma by this current-driven drift-wave
instability, as {\em  experimentally measured}  in \citet{hat}, has
similar properties as the one discussed in the previous text. This
implies the following facts: i)~the anisotropy in the ion heating
($T_{i\bot}> T_{iz}$, ii)~the heating time being comparable to the
growth time of the drift wave and proportional to the wave energy
(square amplitude), iii)~the perpendicular heating being associated
with the leading order perturbed perpendicular drift, while the
parallel heating is associated with the ion Landau damping.

We stress that the {\em collisional} counterpart of the above
described current-driven drift-wave instability, applicable to the
lower solar atmosphere,   may be found in \citet{el} where it has
been explained and {\em also experimentally verified}.

\section{Ion-cyclotron drift wave}

In the case of high-frequency electrostatic perturbations satisfying
the condition $\omega_r \sim\Omega_i$, and in the presence of a
density gradient, we have an ion-cyclotron drift  (ICD) wave, also
called the ion Bernstein mode in the literature \citep{ichi}. This
mode has been theoretically predicted \citep{mih3} and
experimentally verified \citep{hen} long ago. The kinetic
description given in \citet{ichi,mih3,hen} reveals the same
instability mechanism as in the case of the kinetic drift wave
instability presented before. However, in order for the instability
to take place, the drift-wave branch and the ion-cyclotron  branch
must get close to each other. In that case, instead of the
intersection  of two dispersion lines, we have only one
complex-conjugate solution. Clearly, to have this, the equilibrium
density scale length must be very short to make the frequency of the
drift part high, this because for the drift wave $\omega_r\sim
1/L_n$. An application to solar plasma of this instability is
discussed in \citet{v4}. As shown in \citet{ichi}, the ICD mode
grows if the following instability condition is satisfied:
\[
S(k_y, L_n)= 2 k_y^2 \lambda_d^2 \omega_1 (\omega_1 + \omega_2)
\delta
\]
\be
- (\omega_1 - \omega_2)^2 (1+ k_y^2\lambda_d^2  - 2 \delta)>0.
\label{ic1} \ee
Here, $\omega_1= \omega_{*e}/(1+ k_y^2 \lambda_d^2)$, $\omega_2=
\Omega_i \left[1+ \delta/(1+ k_y^2 \lambda_d^2)\right]$,
$\lambda_d^2= \rho_i^2 m_e/m_i + \lambda_{de}^2 T_i/T_e$, $
\lambda_{de}=\vte/\omega_{pe}, \quad \delta=1/[(2 \pi)^{1/2} k_y
\rho_i]$

We stress that the ion-cyclotron mode itself has  been discussed a
lot in the past in the context of  problems related to the heating
of the solar corona, see e.g. \citet{cr1}, \citet{cr2}, \citet{hol}.
The reasons for this are the same as those discussed in the previous
sections: the evidence obtained from in situ measurements in the
solar wind and coronal holes of resonant ion cyclotron heating, and
a preferential heating of coronal ions (with respect to electrons),
that in the same time  is most dominant in the direction
perpendicular to the magnetic field lines. In this context, {\em the
damping} of such IC waves is believed to be a good candidate for the
consequent coronal heating and solar wind acceleration \citep{mark}.
Yet,  as in many wave-heating scenarios of the corona proposed in
the past, there is the problem of the source for the required
generation of such IC waves throughout the corona. The effects
proposed as sources for the IC mode are  global resonant MHD modes
\citep{mark}, currents \citep{for,to},  etc. However, as a rule,
these effects themselves need some source, so the problem is not
solved but merely shifted to another problem.

\begin{figure}
\includegraphics[height=7cm, bb=10 10 300 230,clip=,width=.95\columnwidth]{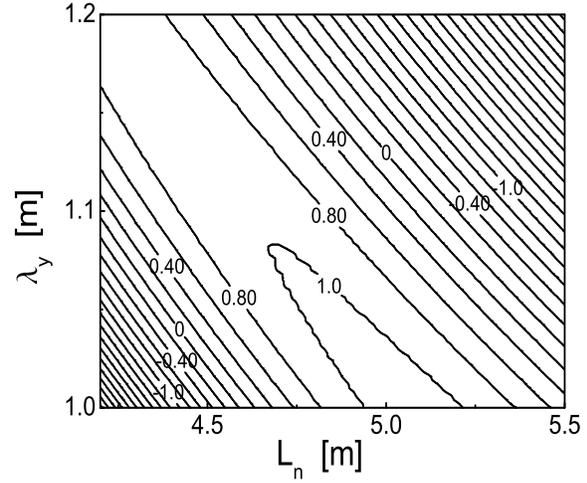}
 \caption{Contour plot of  $S(k_y, L_n)$ from
Eq.~(\ref{ic1}). The ICD mode is growing for  $S(k_y, L_n)>0$.
}\label{fig11}
\end{figure}

The  analysis performed in \citet{v4} and in the previous sections,
demonstrates that such a source (the density gradient) exists, and
the mechanism which it implies is well known in the literature but
not used or studied in the context of the solar plasma, and it is
able to produce growing ICD modes on massive scales. For example, in
Fig.~9 we give the contour plot of $S(k_y, L_n)$ from
Eq.~(\ref{ic1}). The figure shows a narrow range of $L_n$ and
$\lambda_y$ within which the ICD is growing, $S(k_y, L_n)>0$. The
parameters are the same as before: $B_0=10^{-3}\;$T,
$n_0=10^{13}\;$m$^{-3}$, and $T_e=T_i=10^6\;$K. The instability
shown here implies short-scale density inhomogeneities. However,
this scale can be  increased for example by  decreasing  the
magnetic field  intensity. Setting  $10^{-4}\;$T and also reducing
the number density by one order of magnitude, yields the necessary
density scale length $L_n$ for the unstable modes  of the order of
60 meters. Observe that such reduced values may be obtained e.g.\ by
moving further along a flux tube, implying that the mechanism may
work at various altitudes in the solar atmosphere.

\subsection{Stochastic heating of plasma by ICD  wave}

A {\em stochastic} heating mechanism, similar to the case discussed
previously for the drift wave, is also known to exist in this
particular frequency domain. This has been discussed in
\citet{kar2}. It is shown that the stochastic heating by a
perpendicularly propagating electrostatic wave, satisfying the
condition $\omega_r \simeq l \Omega_i$, $l=1, 2, \cdots$,  takes
place if the mode amplitude exceeds the limit:
\be
E>\frac{B_0}{4} \frac{\omega_r}{k_\bot}
\left(\frac{\Omega_i}{\omega_r}\right)^{1/3}.\label{ic3} \ee
The condition is obtained after analyzing the same equations as
Eq.~(\ref{ht1}). This yields the wave potential $\phi=B_0 \Omega_i
l^{2/3}/4=  24 l^{2/3}$. The results from \citet{v4} provide the
maximum mode growth rate $\omega_i/\omega_r\simeq 0.1$. For the
given magnetic field we have $\Omega_i\simeq 10^5\;$Hz for protons.
Now, assuming small starting electrostatic perturbations $e
\phi/(\kappa T_i)\simeq 0.01$, and setting $l=1$, we find out that
the required wave potential amplitude is achieved within about
$3\cdot 10^{-4}\;$s only! Hence, in the given coronal environment
the stochastic heating condition (\ref{ic3}) is practically
instantaneous. It develops in the way already described in the
pervious text.

\section{Concluding comments}

We have presented a few (out of many) physical effects that make the
drift mode growing. There exist many more phenomena that lead to the
growth of drift waves but these are not discussed here at all, e.g.\
the temperature gradient driven drift wave instability, shear flow
instability, etc. As a matter of fact, the presence of a temperature
gradient implies an  additional source of energy and an instability
provided that the temperature and density gradients have opposite
sign \citep{rs}, i.e., $\partial \ln T/\partial \ln n_0 <0$. One
particularly strong mechanism leading to an instability is a sheared
flow. The term `shear flow' refers to the flow of the plasma as a
whole along the magnetic field vector and having in the same time a
gradient in the perpendicular direction. Such a current-less
instability [known also as D'Angelo mode \citep{dasf}] has an extra
energy source in the flow gradient. A kinetic theory analysis of
this instability as given  in \citet{gan}, while its most recent
{\em experimental verification}  may be found in \citet{kan1}. In
\citet{sal} such a shear flow instability is discussed within the
fluid theory in application to coronal spicules, showing strongly
growing modes with the possibility for $\omega_i>\omega_r$. Because
of the radial density gradients, solar spicules are an ideal nursery
for the drift wave instability. Growing drift waves are simply
carried upwards by the plasma flow inside each of them. In the same
time they are numerous (at least $10^5$ of them are present
throughout the Sun at every moment). There are also evidences of the
presence of plasma flows along the magnetic flux loops \citet{sc}.
While the magnitude of these axial flows is determined (showing
subsonic flows with velocities up to $100\;$km/s), their eventual
inhomogeneity in the radial direction remains an open question. Yet,
there are no obvious physical reasons that would exclude them. The
inflow of more dense plasma along a magnetic flux tube from lower
(cooler) layers may then imply the simultaneous presence of a
density gradient (towards the axis) and a temperature gradient
(outward oriented). In such a geometry, the mentioned
temperature-gradient instability may develop, or/and it may be
combined with the shear-flow instability.

In reality, and in particular in the heating processes in the solar
corona, the simple growth-damping process of a coherent  wave, and
the consequent heating due to the resonant  interaction with plasma
particles, can not be a completely accurate picture. It is rather an
interplay between various processes that may happen in the same time
and place, like collisions, mode growth, kinetic effects,
nonlinearity and turbulence, and stochastic ion and electron
heating. The results presented here show that the stochastic heating
related to the drift wave is a powerful mechanism that switches on
for sufficiently large mode amplitudes. The properties of the
resulting heating process are such that they are consistent with all
the observed  features of the plasma heating in the solar
atmosphere. On the other hand, there is plenty of energy  for the
mode itself, stored in the omnipresent density gradients. To our
knowledge, there is no other available heating model like the one
presented in this work, that is so clearly able to fulfill all these
requirements.

The presented mechanism  removes the necessity for explaining the
most crucial problems of coronal heating: namely i)~how the (right
amount of) energy is transmitted from `sources below' the
photosphere to the corona, and ii)~how this energy is dissipated
locally in the corona (at the right rate). Instead, we showed,
first, that a sufficient amount of energy for driving drift modes is
already present in the corona, and, second, that it is naturally
transmitted to the different plasma species by well known effects
that are, however, beyond the standardly used models and theories.
Hence, the proposed mechanism is based on a novel paradigm that
allows a self-consistent solution model. The heating mechanism
implies instabilities on time and spatial scales that are currently
not directly observable by space probes. However, all the effects
presented here are directly experimentally verified under laboratory
conditions.  Their indirect confirmation in the context of the solar
corona seems to be also beyond doubts. This is because the {\em
consequences} of the heating process, as enlisted earlier in the
text (temperature anisotropy, better heating of heavier ions, hotter
ions than electrons), are indeed verified by satellite observations.

\section*{Acknowledgments}

The  results presented here  are  obtained in the framework of the
projects G.0304.07 (FWO-Vlaanderen), C~90347 (Prodex),  GOA/2009-009
(K.U.Leuven). Financial support by the European Commission through
the SOLAIRE Network (MTRN-CT-2006-035484) is gratefully
acknowledged. TRACE (Fig.~1) is a mission of the Stanford-Lockheed
Institute for Space Research, and part of the NASA Small Explorer
program.

\bsp

\label{lastpage}

\end{document}